\documentclass[aps,amsfonts]{revtex4}

\begin{document}
\newcommand{\bb}{\begin{equation}}
\newcommand{\ee}{\end{equation}}
\newcommand{\eqb}{\begin{eqnarray}}
\newcommand{\eqf}{\end{eqnarray}}

\preprint{}
\title{Ultraviolet modified photons and anisotropies in the 
cosmic microwave background radiation}
\author{J.\  Gamboa}
\email{jgamboa@lauca.usach.cl} \affiliation{Departamento de
F{\'\i}sica, Universidad de Santiago de Chile, Casilla 307, Santiago 2,
Chile}
\author{J.\  L\'opez-Sarri\'on}
\email{justo@dftuz.unizar.es} 
\affiliation{Departamento de F{\'\i}sica,
Universidad de Santiago de Chile, Casilla 307, Santiago 2, Chile}
\author{A.\ P.\ Polychronakos}
\email{alexios@sci.ccny.cuny.edu} 
\affiliation{Physics Department, City College of CUNY, New York, NY 10031, USA}

\begin{abstract}
We discuss a minimal canonical modification of electrodynamics in order to account 
for ultraviolet Lorentz violating effects. This modification creates a birefringence
that rotates the polarization planes from
different directions. Such effects might be detectable in the anisotropic 
polarization of the Cosmic Microwave Background radiation.

\end{abstract}
\pacs{PACS numbers:}

\maketitle
The cornerstone of modern cosmology is the cosmological principle, which is
based on the notion that spacetime is locally Lorenz invariant. The analysis,
however, of problems such as
the matter-antimatter asymmetry, the origin of dark matter/energy, or even
the nature of the primordial magnetic 
field, calls for a critical reconsideration of the principles underlying the
cosmological standard model \cite{cosmo}. 

In this direction, several authors have put to the test the validity of Lorentz symmetry \cite{kost}
in the propagation of
light from faraway galaxies \cite{CFJ,ralston,CF}. The theoretical framework of this analysis has mostly been
a Maxwell-Chern-Simons model \cite{CFJ,andrianov}, which introduces a parameter with dimensions of energy
and would represent a correction to electrodynamics at very large scales
and very low energies.

A distinct possibility would be to investigate Lorentz violating effects due to highly
energetic processes in light 
propagation. An excellent candidate to test such a phenomenon would be the
study of anisotropies 
in the Cosmic Microwave Background (CMB) radiation. This is because, even though the mean
temperature of CMB is only
$2.275 K$, it is just a relic of events which happened at the first
epochs of our Universe  --at the decoupling era or much earlier--
where the typical energies were high enough to spur new
ultraviolet effects.\footnote{For a recent discussion about this topic in the context of COBE and WMAP 
see \cite{cobe,wmap}} Furthermore, even a tiny asymmetric effect on the shifting of
polarization planes would be amplified due to the very large
distances that those photons have traveled around to reach us.

In this letter, we will explore the reasonable possibility that in the early epochs
of our Universe, when 
it was mainly dominated by radiation, the electromagnetic processes were not
necessarily described by standard 
relativistic theory. In other words, Maxwell theory should be modified in the
ultraviolet regime. From this point of view our work
would be closer to the one of Myers and Pospelov \cite{MP}.

One approach to modifying Lorentz symmetry at high energies invokes deformations
of the Lorentz group involving an invariant length (of the order of
the Planck length) \cite{Def}. Such deformations are, in fact, nonlinear realizations of the
Lorentz group and they can be mapped to standard Lorentz transformations by a
nonlinear map of the momenta. The single-particle dispersion relations, then, are mapped
to the standard ones under this map. These modified symmetries, however, do not possess a
nontrivial co-product; that is, there is no way to compose two representations
into a new representation of the group, other than the standard one as obtained
by the usual addition of momenta via the nonlinear momentum map. As such, there
are no interacting field theories that realize the modified transformations in a
nontrivial way and therefore no physically interesting effects.

We shall consider, instead, modifying the electromagnetic theory by including small
Lorentz violating terms in the lagrangian. 
The main issue, of course, is what these modifications might be and what are the
criteria for selecting and including relevant and reasonable contributions.
An effective field theory deriving from an underlying fundamental theory (string theory
or other), would involve in principle many possible ultraviolet terms in the
effective action, leaving the question for identifying the relevant ones wide open.

In this work we take the approach of including a minimal modification to the
{\it canonical structure} of the electromagnetic field theory, amounting to adding 
a tiny violation of the microcausality principle. This procedure, proposed in 
\cite{justo,justin},
includes small modifications to the canonical commutators in the Maxwell theory
\eqb
\left[ A_i (x),A_j(y)\right]&=& \epsilon_{ijk}\theta_k \delta (x-y), \label{1}
\\
\left[ \pi_i (x),\pi_j(y)\right]&=& \epsilon_{ijk}\gamma_k \delta (x-y), \label{2}
\\
\left[ A_i (x),\pi_j(y)\right]&=& \delta_{ijk} \delta (x-y), \label{3}
\eqf
where $\theta_i$ and $\gamma_i$ are two given Lorentz violating vectors which 
play the role of ultraviolet and infrared energy scales
respectively.\footnote{In fact, the $|{\vec \theta}|$ and $|{\vec \gamma}|$ respectively.}

If we are interested in ultraviolet effects, we can neglect the infrared
scale $(\gamma\sim 0)$, and retain the corresponding ultraviolet parameter.

The action that reproduces this modified electrodynamics and is consistent with the
commutators (\ref{1})-(\ref{3}) is given by the general form
\bb
S = \int d^4x {\cal L} = \int d^4x \frac{1}{2} \Omega_{ab}\psi^a\dot\psi^b - V(\psi),
\end{equation}
where $\psi^a$ and $\dot\psi^a$ are the coordinates and velocities
with $a=1$, $2$, \dots, $2n$, and $\Omega_{ab}$ is a constant
  antisymmetric and regular matrix. Here, $V$ is a potential which
  modifies the free theory.
From this lagrangian the Poisson structure obtains as 
\begin{equation}
\left\{\psi^a,\psi^b\right\} = (\Omega^{-1})^{ab}.
\end{equation}

We take, therefore, a set of Fields $\{A_i(x), F_j(x) \}$ with $i,j=1,2,3$,
whose Poisson brackets are,
\eqb
\left\{A_i(x),A_j(y)\right\} &=& \epsilon_{ijk}\theta_k\delta^3(x-y), \nonumber
\\
\left\{A_i(x),F_j(y)\right\} &=& \delta_{ij}\delta^3(x-y), \label{com1}
\\
\left\{F_i(x),F_j(y)\right\} &=& 0. \nonumber
\eqf

In the basis $\psi^a(x)= \{A_1, A_2, A_3, F_1, F_2,F_3)\}$, then, $\Omega_{ab}$ becomes 
\begin{equation}
\Omega_{ab}= \left(\begin{array}{ c c c c c c}  
0 & 0&0 & -1&0&0\\
0&0&0 & 0&-1&0\\
0&0&0&0&0&-1\\
1&0&0&0&\theta_3&-\theta_2\\
0&1&0&-\theta_3&0&\theta_1\\
0&0&1&\theta_2&-\theta_1
&0\end{array}\right)\,\delta^3(x-y)
\end{equation}

In terms of these variables the lagrangian can be written as 
\eqb
L &=& \frac{1}{2}\int d^3x\,d^3y\,\Omega_{ab}(x-y)\psi^a\dot\psi^b 
+V(\psi), \label{lag22}
\\
&=& \int d^3x\,\left(F_i\dot A_i + \frac{1}{2}\epsilon_{ijk}\theta_k F_i \dot F_j\right) 
- V(A,F)
\eqf

Let us define $F_i=F_{0i}= - F_{i0} = F^{i0} = - F^{0i}$, 
$F_{ij} = \partial_i A_j -\partial_iA_j= F^{ij}$, and a new
auxiliary variable $A^0=-A_0$. Then we can choose as potential $V$ the following,
\begin{equation}
V(A,F) =\int d^3x\,\left(-\frac{1}{2} F_{0i}F^{0i} - F^{0i}\partial_iA_0
+\frac{1}{4}F_{ij}F^{ij} \right)
\end{equation}
This is the minimal potential that regains the standard
electrodynamics when the $\theta$ parameters vanishes. 
Or equivalently,
\eqb
L &=& \int d^3x\,\biggl(\frac{1}{4}F^{\mu\nu}F_{\mu\nu}-\frac{1}{2}
F^{\mu\nu}(\partial_\mu A_\nu - \partial_\nu A_\mu) \nonumber 
\\
&+&
\frac{1}{2}\epsilon_{ijk} \theta^k F^{0i}\partial_0 F^{0j}\biggr), \nonumber
\eqf

The equations of motion as obtained by Hamilton's principle are
\begin{eqnarray}
&&F_{0i} = \partial_0 A_i - \partial_i A_0 -
\epsilon_{ijk}\theta_j\partial_0 F_{0k}\\
&&F_{ij} =\partial_i A_j -\partial_j A_i\\
&&\partial_\nu F^{\mu\nu} = J^\mu
\end{eqnarray}
where $J^\mu$ represents a matter current coupled to $A_\mu$.

By construction, this theory is gauge invariant if $\partial_\mu
 J^\mu=0$, in the sense that a
 transformation of $A_\mu \rightarrow A_\mu + \partial_\mu\Delta$ for
 any arbitrary function $\Delta$, keeps the action and equations of motion
 invariant. Then, the expression 
of $F$'s in terms of $A$'s is given by,
$$F_{0i} =\left(\frac{1}{I_3
   +\Theta\partial_t}\right)_{ij}(\partial_0A_j-\partial_jA_0)$$
$$F_{ij}= \partial_iA_j-\partial_jA_i$$
where $I_3$ is the $3\times 3$ identity matrix and $\Theta_{ij} =
\epsilon_{ijk}\theta_k$. 
Here it is explicit that in terms of the $A_i$ alone the theory is
nonlocal because
of the nonlocal operator $(I+\Theta\partial_t)^{-1}$ in the equation
of $F_{0i}$ in terms of $A_i$.
 
Let us define, as usual, the magnetic field,
$$B_i = \frac{1}{2}\epsilon_{ijk}F^{ij},$$
and the electric field,
$$E_i = F^{0i}.$$
This is the electric field that couples to matter, according to the
equations of motion. Note, thought, that this is not the usual electric
field as defined in terms of the gauge potential fields. We can define
another electric field, which we will 
call ``old electric field'', as
$$\tilde E_i = -\partial_tA^i -\partial_i A^0 = (\delta_{ij} +
\epsilon_{ijk}\theta_k\partial_t) E_j$$
Then, the equations of motion without matter are, 
\begin{eqnarray}
\partial_i E_i &= &0\\
\dot E_i &=&(\vec\nabla\times\vec B)_i\\
\dot B_i &=&-(\vec\nabla\times\vec{\tilde E})_i
\end{eqnarray}
where the last equation is just the Bianchi identity. This identity
can be read in terms of the electric field, {\it i.e.},
\begin{equation}
\dot B_i = -(\vec\nabla\times\vec E)_i - \vec\theta\cdot\vec\nabla\dot
E_i
\end{equation}
where we have made use of the Gauss law. Then, deriving one of the
equations of motion with respect to time, we get,
\begin{equation}
\partial^2 E_i = - \theta_m\epsilon_{ijk} \partial_m\partial_j \dot
E_k
\end{equation}
where $\partial^2 = \partial_t^2 - \vec\nabla^2$. Expressing $E_i$ in terms
of its Fourier transform,
$$ E_i(x) = \int d^3k\, \varepsilon_i(k) e^{i(\omega t-\vec k\cdot \vec
 x)}$$
with the Gauss law implying that $\vec k\cdot \vec\varepsilon(k) = 0$,
 we obtain
$$\left[(\omega^2 -\vec k^2)\delta_{ij} -i \omega(\vec k\cdot\vec\theta)k_k
 \epsilon_{ijk}\right]\varepsilon_j(k) = 0$$
 Diagonalizing this expression, and taking into account the Gauss law, we
 obtain a dispersion relation with two different modes,
\begin{equation}
\omega^2 - k^2(1 \pm \omega\theta\cos\alpha_{k\theta}) = 0 
\end{equation}
where $\alpha_{k\theta}$ is the angle between $\vec k$ and $\vec
  \theta$, and $\theta$ is the length of $\vec\theta$.
Equivalently,
$$\omega_{\pm} = k\left[\sqrt{1+ \frac{1}{4}(\vec k\cdot\vec\theta)^2} \pm
\frac{1}{2}(\vec k\cdot \vec\theta)\right]$$
This theory, then, presents a birefringence, or Faraday-like rotation effect 
with polarization planes shifted by an amount proportional to $\Delta k\approx
\omega^2\theta\cos\alpha_{k\theta}$. This fact is similar to the one in
the model studied in \cite{CFJ, ralston, CF, andrianov}. However, in
that model a tiny Lorentz symmetry violating parameter affected equally
the whole spectrum, while here the effect is increasingly important
for higher frequencies. This is also unlike Lorenz violation induced by
space noncommutativity, which induces no Faraday-like rotation \cite{NC}.

Though the analysis of experimental data is beyond the scope of our
paper, we think that it would be very interesting to look for the effects
of the above dipole anisotropy in the CMB polarization --above and beyond
the dipole anisotropy due to the relative motion of our
galaxy with respect to the CMB rest frame \cite{Roos}.

The possibility of a tiny dipolar anisotropy at large scale in the
propagation of light through the Universe was pointed 
out by Ralston and Nodland \cite{ralston} who argued this fact by
analyzing data from polarized light coming from far galaxies. 
Carroll and Field \cite{CF} reanalyzed the Ralston and Nodland method using another 
procedure and suggested that, even though observational data are not complete, 
the possibility of such anisotropy cannot be totally ruled out.

As we have pointed out in the introduction, in the analyses of the above works
the authors used a
theoretical model based on a Chern-Simons like coupling as a test for a
possible Lorentz symmetry violation.
In this work, we have considered another possible scenario in which the dipolar
anisotropy arises from short distance effects. Which effect, if any, is backed by
observational data is yet to be discovered.

This work was supported in part by US NSF Grant PHY-0353301, FONDECYT 1050114 and MECESUP-0108.

\end{document}